\documentclass[12pt]{article}
\usepackage[english]{babel}
\usepackage{amsmath,amssymb,dsfont, float, bm}
\usepackage{amsmath,amssymb,amsfonts,amsthm,mathtools,mathrsfs,booktabs}
\usepackage{tikz,float}
\usepackage{pgfplots}
\pgfplotsset{compat=1.16}
\usepackage{authblk}

\usepackage{natbib}

\usepackage[colorlinks=true, citecolor=blue, urlcolor=blue]{hyperref}
\usepackage{newtxtext,newtxmath} 
\usepackage{mhchem} 
\usepackage{multirow}
\usepackage{bookmark}
\usepackage{booktabs}
\usepackage{threeparttable}
\usepackage{booktabs}

\hypersetup{                 
colorlinks=true,
linkcolor=blue,
citecolor=blue,
anchorcolor=blue,
urlcolor=blue,
linktoc=page,
}
\usepackage[a4paper,margin=2.5cm]{geometry}

\newtheorem{definition}{Definition}[section]

\newtheorem{proposition}{Proposition}[section]

\newcommand{\KS}{d_{\mathrm{KS}}}

\begin{document}

\title{\bf A rank-associated measure of population heterogeneity based on order statistics}

\author[]{Roberto Vila }
 \author[]{\, Frederico Almeida \thanks{Corresponding author: frederico.almeida@unb.br }}
\affil[]{Department of Statistics, University of
	Bras\'ilia, 70910-900, Bras\'ilia, Brazil}

\setcounter{Maxaffil}{0}
\renewcommand\Affilfont{\itshape\small}


\maketitle
\begin{abstract}

This paper introduces a rank-associated measure of population heterogeneity that quantifies how population differences are reflected in the relative positions of observations within a pooled sample. The measure is defined as a weighted average of Kolmogorov–Smirnov distances between conditional and unconditional distributions of order statistics, conditional on the population of origin at each rank. We establish its main properties, including boundedness, invariance under common strictly monotone transformations, and a characterization of its null value through independence between order statistics and population indicators. We derive bounds relating the measure to weighted pairwise distributional distances and population diversity across ranks, and examine the contributions of the minimum and maximum, obtaining a simple expression for two populations. Monte Carlo simulations with four populations assess its finite-sample behavior and asymptotic convergence. The empirical application reveals a shift in heterogeneity toward higher ranks in the germ-free group.

\end{abstract}
	\smallskip
	\noindent
	{\small {\bfseries Keywords.} {
    Heterogeneity; order statistics; rank indicators;
    conditional distributions; Kolmogorov--Smirnov distance; independent
    populations.
}}
	\\
	{\small{\bfseries Mathematics Subject Classification (2020).} 60E05, 60E15, 62N01, 62G30, 62E10.}


\section{Introduction}
\label{Introduction}

Comparing heterogeneous populations is an important problem in statistics, with applications in reliability and survival analysis, economics, environmental studies, and experimental research. A common approach is to compare their marginal distributions. For continuous distributions $F_1,\ldots,F_m$, one possibility is the average Kolmogorov--Smirnov distance,
\[
M_m
\coloneqq
\frac{2}{m(m-1)}
\sum_{1\leqslant i<j\leqslant m}
\KS(F_i,F_j),
\]
where
$
\KS(F,G)=\sup_{x\in\mathbb R}|F(x)-G(x)|
$
denotes the Kolmogorov--Smirnov distance \citep{Massey1951,GibbsSu2002}.
Although useful for comparing marginal distributions, such measures do not account for how population identities are reflected in the relative positions of observations in the pooled sample.

In this paper, we approach population heterogeneity from this perspective. Consider independent random variables $X_1,\ldots,X_m$, where $X_i$ represents an observation from population $i$, and let $X_{1:m}\leqslant\cdots\leqslant X_{m:m}$ denote their order statistics \citep{DavidNagaraja2003}. The indicator
$
\delta_{k,i}=\mathds{1}_{\{X_i=X_{k:m}\}}
$
identifies whether the observation from population $i$ occupies rank $k$. The relationship between order statistics and their associated rank indicators has been studied in several contexts. \cite{BapatKochar1993} established characterizations under proportional hazard rates, while \cite{Malov1996} investigated related results through order structures. \cite{Kella(2024)} considered the independence between time and cause. More recently, \citet{VilaAlmeida2026} developed general characterizations of independence between order statistics and rank indicators, with particular attention to the minimum and maximum in right- and left-censoring models, respectively.

Motivated by this line of research, we introduce a rank-associated measure of population heterogeneity that considers all order statistics and population indicators. The measure is constructed by comparing the conditional distribution of each order statistic, given the population supplying the corresponding observation, with its unconditional distribution. We aggregate the resulting Kolmogorov--Smirnov distances across ranks and populations, weighting each contribution by $p_{k,i}=\mathbb{P}(\delta_{k,i}=1)$. In this way, the measure captures how population identities are distributed across the pooled ordering, complementing comparisons based solely on marginal distributions.

We establish the main properties of the proposed measure, including boundedness, invariance under common strictly monotone transformations, and a characterization of its null value in terms of independence between order statistics and population indicators. We also derive bounds connecting it to weighted pairwise distributional distances and to the diversity of population identities across ranks. Finally, we examine its decomposition into contributions from the minimum, maximum, and intermediate order statistics.

The remainder of the paper is organized as follows. Section \ref{A rank-associated measure of heterogeneity} introduces the measure and establishes its main properties. Section \ref{The case of two populations} considers two populations through an example involving independent exponential random variables. Section \ref{Comparison with marginal heterogeneity} compares the proposed measure with marginal heterogeneity using a lognormal example. Section \ref{simu-and-application} presents simulation results and an application to real data. Finally, Section \ref{Discussion} discusses the interpretation and limitations of the proposed measure, summarizes the main
findings, and outlines directions for future research.

\section{A rank-associated measure of heterogeneity}\label{A rank-associated measure of heterogeneity}

Let $X_1,\ldots,X_m$ be independent, absolutely continuous random variables, with respective distribution functions
$F_1,\ldots,F_m$ and densities $f_1,\ldots,f_m$. Throughout this section, assume that ties occur with probability zero;
that is, $\mathbb{P}(X_i=X_j)=0$ for all $i\neq j$.

For $k,i\in\{1,\ldots,m\}$, define the rank indicator
$
\delta_{k,i}
\coloneqq
\mathds{1}_{\{X_i=X_{k:m}\}},
$
which equals one if the observation from population $i$ occupies
rank $k$ and zero otherwise. Let
\begin{align}\label{def-p}
p_{k,i}=\mathbb{P}(\delta_{k,i}=1).
\end{align}
For each rank $k$, these probabilities satisfy
$
\sum_{i=1}^m p_{k,i}=1.
$
Moreover, for each population $i$,
$
\sum_{k=1}^m p_{k,i}=1.
$

Whenever $p_{k,i}>0$, define the conditional distribution function
\begin{equation}\label{def-Fki}
	F_{k,i}^*(x)
	\coloneqq
	\mathbb{P}(X_i\leqslant x\mid \delta_{k,i}=1).
	\end{equation}
	This is the distribution of an observation from population $i$
	conditional on its occupying rank $k$ in the pooled ordering.
	This conditional representation is closely related to the
	characterizations of independence between order statistics and
	rank indicators considered in \citet{VilaAlmeida2026}.
	
	The distribution function of the $k$th order statistic is denoted
	by $F_{k:m}$. Since exactly one population supplies the observation
	occupying rank $k$, the law of total probability gives
	\begin{equation}
		F_{k:m}(x)
		=
		\sum_{i=1}^m p_{k,i}F_{k,i}^*(x),
		\label{eq:mixture}
	\end{equation}
	where $p_{k,i}$ is as given in \eqref{def-p}.
	
	Thus, the distribution of an order statistic is a mixture of the
	conditional distributions associated with the possible population
	identities, with mixing probabilities given by the rank probabilities.
	
	\begin{definition}
		The rank-associated heterogeneity measure is defined as
		\begin{align}
			H_m
			\coloneqq
			H_m
			(X_1,\ldots,X_m)
			=
			\frac{1}{m}
			\sum_{k=1}^m
			\sum_{i=1}^m
			p_{k,i}
			\KS(F_{k,i}^*,F_{k:m}),
			\label{eq:Hm}
		\end{align}
		where terms with $p_{k,i}=0$ are defined to be zero.
	\end{definition}
	
	The measure averages, over all ranks, the discrepancies between
	the conditional distribution associated with each population and
	the unconditional distribution of the corresponding order statistic.
	The factor $1/m$ gives equal weight to the different ranks, while
	$p_{k,i}$ reflects the probability that population $i$ occupies
	rank $k$.
	
	The measure has a direct probabilistic interpretation. A large
	value of $H_m$ indicates that knowing the population identity of
	an observation occupying a particular rank can substantially change
	the distribution of the corresponding order statistic. A small value
	indicates that the conditional distributions are close to their
	unconditional counterparts.
	
	\begin{proposition}
		The measure $H_m$ satisfies
		$
		0\leqslant H_m\leqslant 1.
		$
		
		Furthermore, $H_m=0$ if and only if
		$
		F_{k,i}^*=F_{k:m}
		$
		for every pair $(k,i)$ such that $p_{k,i}>0$.
	\end{proposition}
	
	\begin{proof}
		Since the Kolmogorov--Smirnov distance between two distribution
		functions belongs to $[0,1]$, and since
		$
		\sum_{i=1}^m p_{k,i}=1,
		$
		we obtain
		\begin{equation*}
		0\leqslant
		\sum_{i=1}^m p_{k,i}
		\KS(F_{k,i}^*,F_{k:m})
		\leqslant 1.
	\end{equation*}
	Averaging over the $m$ ranks yields $0\leqslant H_m\leqslant 1$.
	
	Since all terms in \eqref{eq:Hm} are nonnegative, $H_m=0$ if and only if
	every term with positive weight is zero. This is equivalent to
	the stated equality of distribution functions.
\end{proof}

The preceding result establishes that the proposed measure is zero
precisely when the population identity does not affect the
distribution of the order statistic, conditional on the rank being
specified.

This interpretation can be expressed in terms of independence.

\begin{proposition}
	For any $k$ and $i$ such that $p_{k,i}>0$,
	$X_{k:m}$ and $\delta_{k,i}$ are independent if and only if
	$
	F_{k,i}^*=F_{k:m}.
	$
	
	Consequently, $H_m=0$ if and only if $X_{k:m}$ is independent of
	$\delta_{k,i}$ for every pair $(k,i)$ with $p_{k,i}>0$.
\end{proposition}
\begin{proof}
	The first equivalence follows from Theorem 1 of
	\citet{VilaAlmeida2026}, which establishes that the independence
	between an order statistic and its corresponding rank indicator
	is equivalent to the equality of the associated conditional and
	unconditional distributions.
	
	For the second assertion, since all terms defining $H_m$ are
	nonnegative, $H_m=0$ if and only if
	$
	F_{k,i}^*=F_{k:m}
	$
	for every pair $(k,i)$ such that $p_{k,i}>0$.
	The result then follows directly from the first equivalence.
\end{proof}

An immediate consequence is that $H_m=0$ when the random variables
are independent and identically distributed (i.i.d.) with a continuous
distribution. In this case, the population identity is independent of
the rank occupied by the observation, so the conditional distribution
of the corresponding order statistic is the same regardless of the
population from which the observation originates.

\begin{proposition}\label{prop-s}
	For each $k=1,\ldots,m$, let
	\[
	D_k\coloneqq 1-\sum_{i=1}^m p_{k,i}^2.
	\]
	Then
	\[
	H_m
	\leqslant 
	\frac{1}{m}\sum_{k=1}^m D_k
	=
	1-\frac{1}{m}\sum_{k=1}^m\sum_{i=1}^m p_{k,i}^2.
	\]
	In particular,
	$
	H_m\leqslant 1-{1}/{m}.
	$
\end{proposition}
\begin{proof}
	For each $k,i$, the mixture representation \eqref{eq:mixture}
	and the triangle inequality imply
	\begin{align*}
		d_{\mathrm{KS}}(F_{k,i}^*,F_{k:m})
		\leqslant
		\sum_{j=1}^m p_{k,j}
		d_{\mathrm{KS}}(F_{k,i}^*,F_{k,j}^*)
		\leqslant
		\sum_{j\ne i}p_{k,j}
		=1-p_{k,i}.
	\end{align*}
	Consequently,
	\begin{align*}
		H_m
		=\frac{1}{m}\sum_{k=1}^m
		\sum_{i=1}^m p_{k,i}
		d_{\mathrm{KS}}(F_{k,i}^*,F_{k:m})
		\leqslant
		\frac{1}{m}\sum_{k=1}^m
		\sum_{i=1}^m p_{k,i}(1-p_{k,i})
		=
		\frac{1}{m}\sum_{k=1}^m
		\left(1-\sum_{i=1}^m p_{k,i}^2\right).
	\end{align*}
Finally, applying the Cauchy--Schwarz inequality, we have
$
\left(\sum_{i=1}^m p_{k,i}\right)^2
\leqslant
m\sum_{i=1}^m p_{k,i}^2.
$
Since $\sum_{i=1}^m p_{k,i}=1$, this immediately implies
$
\sum_{i=1}^m p_{k,i}^2\geqslant {1}/{m},
$
thereby establishing the stated universal bound.
\end{proof}

	The quantity $D_k$ in Proposition \ref{prop-s}
	is the Gini--Simpson diversity index \citep{Jost2006} associated with the
	population identity of the observation occupying the $k$th
	rank. Thus, the preceding result shows that the contribution
	of each rank to $H_m$ is bounded by the diversity of the
	corresponding rank assignment. In particular, the upper bound
	depends not only on the number of populations but also on
	how evenly the populations are represented at each rank.

\subsection{A representation in terms of the marginal distributions}

The conditional distributions in \eqref{eq:Hm} can be expressed directly
in terms of the marginal distribution functions and densities.

For each $i$, define
\[
W_{k,i}(x)
\coloneqq
\mathbb{P}\left(
\sum_{j\ne i}\mathds{1}_{\{X_j<x\}}=k-1
\right).
\]
This is the probability that exactly $k-1$ observations from
populations other than $i$ are smaller than $x$. 
Since
\[
\left\{
\sum_{j\ne i}\mathds{1}_{\{X_j<x\}}=k-1
\right\}
=
\bigcup_{\substack{
		A\subseteq\{1,\ldots,m\}\setminus\{i\}\\
		|A|=k-1
}}
\left[
\left(\bigcap_{j\in A}\{X_j<x\}\right)
\cap
\left(\bigcap_{\substack{j\notin A\\j\ne i}}
\{X_j\geqslant x\}\right)
\right],
\]
by independence, the probability $W_{k,i}(x)$ can be written as 
\begin{equation*}
	W_{k,i}(x)
	=
	\sum_{\substack{
			A\subseteq\{1,\ldots,m\}\setminus\{i\}\\
			|A|=k-1
	}}
	\prod_{j\in A}F_j(x)
	\prod_{\substack{
			j\notin A\\j\ne i
	}}
	[1-F_j(x)].
\end{equation*}


{
By definition, $\delta_{k,i}=\mathds{1}_{\{X_i=X_{k:m}\}}$. Since the random variables are independent and continuously distributed, conditioning on $X_i=x$ gives
\[
\begin{aligned}
	\mathbb{P}(\delta_{k,i}=1\mid X_i=x)
	=\mathbb{P}(X_i=X_{k:m}\mid X_i=x)
	=\mathbb{P}\left(
	\sum_{j\ne i}\mathds{1}_{\{X_j<x\}}=k-1
	\right)
	=W_{k,i}(x).
\end{aligned}
\]
Consequently, for any measurable set $A\subseteq\mathbb{R}$, the law of total probability yields
\[
\begin{aligned}
	\mathbb{P}(X_i\in A,\delta_{k,i}=1)
	=\int_A
	\mathbb{P}(\delta_{k,i}=1\mid X_i=x)
	f_i(x)\,\mathrm{d}x
	=\int_A f_i(x)W_{k,i}(x)\,\mathrm{d}x.
\end{aligned}
\]
Taking $A=\mathbb{R}$, we obtain the probability that $X_i$ occupies rank $k$:
\[
\begin{aligned}
	p_{k,i}
	\stackrel{\eqref{def-p}}{=}
	\mathbb{P}(\delta_{k,i}=1)
	=
	\int_{-\infty}^{\infty}
	f_i(x)W_{k,i}(x)\,\mathrm{d}x.
\end{aligned}
\]
Furthermore, provided that $p_{k,i}>0$, the conditional distribution of $X_i$ given $\delta_{k,i}=1$ is absolutely continuous, with density
\begin{align}\label{density-*}
	f_{k,i}^*(x)
	=
	f_{X_i\mid\delta_{k,i}=1}(x)
	=
	\frac{f_i(x)W_{k,i}(x)}{p_{k,i}}.
\end{align}
Thus, the rank probabilities 
are given by
\begin{equation}
	\begin{aligned}
		p_{k,i}
		=
		\int_{-\infty}^{\infty}
		f_i(x)W_{k,i}(x)\,\mathrm{d}x.
	\end{aligned}
	\label{eq:pki}
\end{equation}
}

By combining \eqref{eq:mixture} and \eqref{density-*}, the density of the $k$th order statistic is given by
\begin{equation}
	f_{k:m}(x)
	=
	\sum_{i=1}^m f_i(x)W_{k,i}(x).
	\label{eq:order-density}
\end{equation}

The expressions in \eqref{eq:pki} and \eqref{eq:order-density} follow from the standard density representation
of order statistics for independent, not necessarily identically
distributed, random variables; see
\citet{DavidNagaraja2003}.

For the minimum and maximum, these expressions simplify
considerably. In particular,
$
W_{1,i}(x)
=
\prod_{j\ne i}[1-F_j(x)]
$
and
$
W_{m,i}(x)
=
\prod_{j\ne i}F_j(x).
$
It follows that
\begin{align}\label{densities-m-m}
f_{1,i}^*(x)
=
\frac{
	f_i(x)\prod_{j\ne i}[1-F_j(x)]
}{
	p_{1,i}
},
\quad
\text{whereas}
\quad
f_{m,i}^*(x)
=
\frac{
	f_i(x)\prod_{j\ne i}F_j(x)
}{
	p_{m,i}
}.
\end{align}

These two expressions reveal an important distinction between
the lower and upper extremes. For the minimum, the observation
from population $i$ must be smaller than all the remaining
observations. For the maximum, it must be larger than all
the remaining observations. Thus, the conditional distributions
associated with the two extremes reflect different aspects of
the marginal distributions, particularly their lower- and
upper-tail behavior.

\subsection{Connection with pairwise distributional dispersion}

The proposed measure $H_m$ in \eqref{eq:Hm} can also be related to a weighted average
of pairwise distances between the conditional distributions.

For each rank $k$, define
\[
A_{k,m}
\coloneqq
\sum_{i=1}^m
p_{k,i}\KS(F_{k,i}^*,F_{k:m})
\quad
\text{and}
\quad
P_{k,m}
\coloneqq
\sum_{i=1}^m\sum_{j=1}^m
p_{k,i}p_{k,j}
\KS(F_{k,i}^*,F_{k,j}^*),
\]
where $p_{k,i}$ is as given in \eqref{def-p}.

Here, $A_{k,m}$ measures the weighted dispersion of the conditional
distributions around their mixture, whereas $P_{k,m}$ measures
their weighted pairwise dispersion.

\begin{proposition}
	For every $k$,
	\begin{equation}
		\frac{1}{2} \, P_{k,m}
		\leqslant A_{k,m}
		\leqslant P_{k,m}.
		\label{eq:pairwise-bounds}
	\end{equation}
	
	Consequently, defining
	$$
	P_m\coloneqq\frac{1}{m} \sum_{k=1}^m P_{k,m},
	$$
	we have
	\begin{equation}
		\frac{1}{2} \, P_m
		\leqslant H_m
		\leqslant P_m.
		\label{eq:global-pairwise-bounds}
	\end{equation}
\end{proposition}

\begin{proof}
	By the mixture representation \eqref{eq:mixture} and the triangle inequality,
	\[
	\begin{aligned}
		\KS(F_{k,i}^*,F_{k:m})
		=
		\KS\left(
		F_{k,i}^*,
		\sum_{j=1}^m p_{k,j}F_{k,j}^*
		\right)
		\leqslant
		\sum_{j=1}^m p_{k,j}
		\KS(F_{k,i}^*,F_{k,j}^*).
	\end{aligned}
	\]
	Multiplying by $p_{k,i}$ and summing over $i$ gives the upper bound.
	
	For the lower bound, the triangle inequality yields
	\[
	\KS(F_{k,i}^*,F_{k,j}^*)
	\leqslant
	\KS(F_{k,i}^*,F_{k:m})
	+
	\KS(F_{k:m},F_{k,j}^*).
	\]
	Multiplying by $p_{k,i}p_{k,j}$ and summing over $i$ and $j$, we obtain
	$
	P_{k,m}\leqslant 2A_{k,m}.
	$
	This proves \eqref{eq:pairwise-bounds}. Averaging over the ranks gives
	\eqref{eq:global-pairwise-bounds}.
\end{proof}

This result establishes a direct mathematical connection between
the proposed measure and a weighted pairwise distributional
dispersion. The distinction is that $H_m$ measures deviations
from the natural mixture distribution $F_{k:m}$, rather than
comparing all conditional distributions directly.

\subsection{Heterogeneity associated with the minimum and maximum}

The proposed measure $H_m$ in \eqref{eq:Hm} aggregates the contributions of all order
statistics. However, the minimum and maximum deserve particular
attention because they represent the two extreme positions in
the pooled ordering.

Define
\[
H_m^{\min}
\coloneqq
\sum_{i=1}^m
p_{1,i}
\KS(F_{1,i}^*,F_{1:m})
\quad
\text{and}
\quad 
H_m^{\max}
\coloneqq
\sum_{i=1}^m
p_{m,i}
\KS(F_{m,i}^*,F_{m:m}),
\]
where $p_{k,i}$ is as given in \eqref{def-p}.

The contribution of the intermediate order statistics is
\begin{equation*}
	H_m^{\mathrm{int}}
	\coloneqq
	\sum_{k=2}^{m-1}
	\sum_{i=1}^m
	p_{k,i}
	\KS(F_{k,i}^*,F_{k:m}),
\end{equation*}
where $H_m^{\mathrm{int}}=0$ when $m=2$. The overall measure can
then be decomposed as
\begin{equation}
	H_m
	=
	\frac{1}{m}
	\big(
	H_m^{\min}
	+
	H_m^{\max}
	+
	H_m^{\mathrm{int}}
	\big).
	\label{eq:decomposition}
\end{equation}
The decomposition in \eqref{eq:decomposition} makes it possible to distinguish heterogeneity
associated with the lower extreme from heterogeneity associated
with the upper extreme. In particular, $H_m^{\min}$ measures
the extent to which the distribution of the minimum depends on
the population supplying that minimum, whereas $H_m^{\max}$
measures the corresponding dependence for the maximum.

For two populations, there are no intermediate order statistics,
and therefore
\begin{equation}
	H_2
	=
	\frac{1}{2}
	\big(
	H_2^{\min}+H_2^{\max}
	\big).
	\label{eq:H2-decomposition}
\end{equation}

The decomposition in \eqref{eq:H2-decomposition} has two components that also admit a simple pairwise representation.
Let
$
q=\mathbb{P}(X_1<X_2)
$.
%
Since, by \eqref{eq:mixture},
$
F_{1:2}
=
qF_{1,1}^*+(1-q)F_{1,2}^*
\ 
\text{and}
\ 
F_{2:2}
=
(1-q)F_{2,1}^*+qF_{2,2}^*,
$
we obtain
$
H_2^{\min}=2q(1-q)\KS(F_{1,1}^*,F_{1,2}^*)
\
\text{and}
\
H_2^{\max}=2q(1-q)\KS(F_{2,1}^*,F_{2,2}^*).
$
Therefore,
\begin{equation}
	H_2=q(1-q)\big[\KS(F_{1,1}^*,F_{1,2}^*)+\KS(F_{2,1}^*,F_{2,2}^*)\big].
	\label{eq:H2-two-pop}
\end{equation}

The representation in \eqref{eq:H2-two-pop} shows explicitly that the overall
heterogeneity measure for two populations is determined by
the discrepancies between the conditional distributions of
the minimum and maximum, weighted by the probability of the
two possible population orderings.

The relative contribution of the two extremes can also be
summarized by
\begin{equation}
	A_m
	\coloneqq
	\frac{H_m^{\max}-H_m^{\min}}
	{H_m^{\max}+H_m^{\min}},
	\label{eq:asymmetry}
\end{equation}
provided that $H_m^{\max}+H_m^{\min}>0$. It follows from \eqref{eq:asymmetry} that
$
-1\leqslant A_m\leqslant 1.
$

Positive values indicate that the maximum contributes more
than the minimum, whereas negative values indicate the opposite.
A value of zero means that the two extreme contributions are
equal. This quantity complements $H_m$ by describing the
direction of the heterogeneity associated with the extremes.

\subsection{Invariance properties}

The proposed measure $H_m$ in \eqref{eq:Hm} is invariant under a common strictly
increasing transformation of all the observations.

\begin{proposition}
Let $X_1,\ldots,X_m$ be independent, absolutely continuous random
variables, and let $g$ be a strictly monotone function.
	Then
	\[
	H_m(g(X_1),\ldots,g(X_m))
	=
	H_m(X_1,\ldots,X_m).
	\]
	Moreover, if $g$ is strictly decreasing, then
	\[
	H_{k,m}(g(X_1),\ldots,g(X_m))
	=
	H_{m-k+1,m}(X_1,\ldots,X_m),
	\]
	for $k=1,\ldots,m$.
\end{proposition}
\begin{proof}
	If $g$ is strictly increasing, the result follows from
	the invariance of the Kolmogorov--Smirnov distance under
	strictly increasing transformations and the preservation
	of the ranks.
	
	Suppose now that $g$ is strictly decreasing. In this case,
	the ordering of the observations is reversed. In particular,
	the observation occupying rank $k$ after transformation
	is the one that occupied rank $m-k+1$ before transformation.
	Thus, the rank probabilities satisfy
	$
	p_{k,i}^{(g)}
	=
	p_{m-k+1,i}.
	$
	
	Furthermore, the conditional distributions associated
	with rank $k$ after transformation are the images, under
	$g$, of the conditional distributions associated with
	rank $m-k+1$ before transformation. Since the
	Kolmogorov--Smirnov distance is invariant under a common
	strictly monotone transformation, we obtain
	\[
	H_{k,m}(g(X_1),\ldots,g(X_m))
	=
	H_{m-k+1,m}(X_1,\ldots,X_m).
	\]
	Summing over $k$ and using the fact that
	$k\mapsto m-k+1$ is a permutation of
	$\{1,\ldots,m\}$ gives
	\[
	H_m(g(X_1),\ldots,g(X_m))
	=
	H_m(X_1,\ldots,X_m).
	\]
	This completes the proof.
\end{proof}

	The preceding result shows that the proposed measure is
	invariant under both increasing and decreasing changes
	of scale. Although an increasing transformation preserves
	the ordering of the observations, a decreasing
	transformation reverses the rank-specific contributions.
	Consequently, the overall heterogeneity measure remains
	unchanged, whereas its rank profile is reflected around
	the central rank.

\section{The case of two populations}\label{The case of two populations}

The case $m=2$ provides a particularly transparent interpretation
of the proposed measure.

Let $X_1$ and $X_2$ be independent, absolutely continuous random
variables with distribution functions $F_1$ and $F_2$,
respectively. Define
$
q=\mathbb{P}(X_1<X_2).
$
Then
$
p_{1,1}=q,\ p_{1,2}=1-q,
$
and
$
p_{2,1}=1-q,\ p_{2,2}=q.
$

From \eqref{densities-m-m} the conditional densities associated with the minimum are
\begin{align}\label{pdf-1}
f_{1,1}^*(x)
=
\frac{f_1(x)[1-F_2(x)]}{q},
\quad
f_{1,2}^*(x)
=
\frac{f_2(x)[1-F_1(x)]}{1-q},
\end{align}
and those associated with the maximum are
\begin{align}\label{pdf-2}
f_{2,1}^*(x)
=
\frac{f_1(x)F_2(x)}{1-q},
\quad
f_{2,2}^*(x)
=
\frac{f_2(x)F_1(x)}{q}.
\end{align}

These expressions are useful for both theoretical investigations
and numerical evaluation; see also \citet{DavidNagaraja2003}.

\subsection{An illustrative example: exponential populations}

The distinction between the minimum and maximum contributions
becomes particularly clear when the two populations follow
exponential distributions. Let
$
X_1\sim\operatorname{exp}(\theta_1),
\
X_2\sim\operatorname{exp}(\theta_2),
$
independently, where $\theta_1,\theta_2>0$ are the rate parameters.

The probability that population 1 supplies the minimum is
\[
q=\mathbb{P}(X_1<X_2)
=
\frac{\theta_1}{\theta_1+\theta_2}.
\]

The conditional distribution of the minimum given that it comes
from population 1 is exponential with rate $\theta_1+\theta_2$.
The same distribution is obtained when the minimum comes from
population 2. Hence,
$
F_{1,1}^*=F_{1,2}^*,
$
which implies
\begin{equation}
	\KS(F_{1,1}^*,F_{1,2}^*)=0
	\quad\text{and}\quad
	H_2^{\min}=0.
	\label{eq:exp-min}
\end{equation}

The situation is different for the maximum. Its conditional
survival functions are
\[
S_{2,1}^*(x)
=
\frac{\theta_1+\theta_2}{\theta_2} \,
e^{-\theta_1x}
-
\frac{\theta_1}{\theta_2} \,
e^{-(\theta_1+\theta_2)x}
\quad
\text{and}
\quad
S_{2,2}^*(x)
=
\frac{\theta_1+\theta_2}{\theta_1} \,
e^{-\theta_2x}
-
\frac{\theta_2}{\theta_1} \,
e^{-(\theta_1+\theta_2)x}.
\]

These distributions are different whenever
$\theta_1\ne\theta_2$. Therefore,
\[
H_2^{\max}
=
2q(1-q)
\KS(F_{2,1}^*,F_{2,2}^*)
>0
\]
for unequal rates. Combining this result with
\eqref{eq:exp-min}, we obtain
\begin{equation}
	H_2^{\min}=0,\quad
	H_2^{\max}>0
	\quad\text{if }\theta_1\ne\theta_2.
	\label{eq:exp-extremes}
\end{equation}

Thus, by \eqref{eq:exp-extremes}, for two exponential populations with different rates,
the proposed measure detects heterogeneity through the maximum
but not through the minimum. This behavior is a consequence of
the memoryless property of the exponential distribution:
conditional on being the first observation, the minimum has
the same exponential distribution regardless of which population
supplies it; see \citet{Ross2014}.

This example illustrates why separating the minimum and maximum
contributions can be informative. Although both extremes are
derived from the same two marginal distributions, their
conditional distributions may exhibit markedly different
behavior. In this case, the minimum does not reveal the difference
between the populations through the proposed conditional
comparison, whereas the maximum does.

\section{Comparison with marginal heterogeneity}\label{Comparison with marginal heterogeneity}

We now investigate whether the proposed measure $H_m$ in \eqref{eq:Hm}  provides
information that is not captured by a conventional marginal
heterogeneity measure.

For two populations, consider the marginal Kolmogorov--Smirnov
measure
\begin{equation}
	M_2=\KS(F_1,F_2).
	\label{eq:M2}
\end{equation}

Unlike $H_2$, the measure in \eqref{eq:M2} compares the original marginal
distributions directly. It does not explicitly account for
the conditional distributions associated with the minimum
and maximum.

The following example demonstrates that two population
configurations can have the same value of $M_2$ but substantially
different values of $H_2$.

Let $\operatorname{LN}(\mu,\sigma^2)$ denote the lognormal
distribution with parameters $\mu$ and $\sigma^2$, so that
its logarithm follows a normal distribution with mean $\mu$
and variance $\sigma^2$.

Consider the following two configurations.

\paragraph{Configuration A.}
$
X_1\sim\operatorname{LN}(0,0.5^2),
\ 
X_2\sim\operatorname{LN}(0.49585,0.5^2).
$

\paragraph{Configuration B.}
$
X_1\sim\operatorname{LN}(0,0.5^2),
\ 
X_2\sim\operatorname{LN}(0.54935,1.5^2).
$

In both configurations, the populations are independent.
The parameters of the second population were selected so that
the marginal Kolmogorov--Smirnov distances are approximately
equal. By using densities in \eqref{pdf-1} and \eqref{pdf-2}, numerical integration gives the results reported in
Table \ref{Table-1}.
\begin{table}[H]
	\centering
	\caption{Marginal and rank-associated heterogeneity measures
		for the two lognormal configurations.}
	\label{tab:lognormal}
	\begin{tabular}{lcc}
		\toprule
		Quantity & Configuration A & Configuration B\\
		\midrule
		$M_2$ & 0.3800 & 0.3800\\
		$q=\mathbb{P}(X_1<X_2)$ & 0.7584 & 0.6359\\
		$\KS(F_{1,1}^*,F_{1,2}^*)$ & 0.1758 & 0.4469\\
		$\KS(F_{2,1}^*,F_{2,2}^*)$ & 0.1758 & 0.5850\\
		$H_2$ & 0.0644 & 0.2389\\
		\bottomrule
	\end{tabular}
	\label{Table-1}
\end{table}
%
The values of $H_2$ are obtained from \eqref{eq:H2-two-pop}. 
For Configuration A,
$
H_2
\approx
(0.7584)(0.2416)(0.1758+0.1758)
\approx 0.0644.
$
For Configuration B,
$
H_2
\approx
(0.6359)(0.3641)(0.4469+0.5850)
\approx 0.2389.
$

Thus, although the two configurations have essentially the
same marginal heterogeneity, their rank-associated heterogeneity
measures differ considerably. In fact,
$
{H_2^{(B)}}/{H_2^{(A)}}\approx 3.71.
$

This difference illustrates the additional information provided
by the proposed measure.

In Configuration A, the two populations have the same log-scale
standard deviation. Their difference is primarily associated
with a shift in location. The conditional distributions associated
with the minimum and maximum are relatively close.
In Configuration B, the populations differ in both location
and scale. In particular, the second population has a substantially
larger log-scale standard deviation. Although the marginal
Kolmogorov--Smirnov distance remains approximately the same as
in Configuration A, the conditional distributions associated
with the ranks are much more different.

The example highlights an important distinction. The marginal
measure $M_2$ quantifies how different the two original
distributions are, whereas $H_2$ quantifies how different
the rank-conditioned distributions are from the corresponding
unconditional order-statistic distributions. The two measures
therefore describe related but distinct aspects of population
heterogeneity.

The separate contributions of the minimum and maximum also
provide useful information. The two configurations can be
compared not only through the overall value of $H_2$, but also
through the relative importance of $H_2^{\min}$ and
$H_2^{\max}$. This distinction helps identify whether the
observed rank-associated heterogeneity is primarily related
to the lower or upper extreme.

\section{Simulation results and application}\label{simu-and-application}

{
This section presents a numerical evaluation of the asymptotic and finite-sample properties of the proposed heterogeneity measures, using a comprehensive simulation study and an application to real data. Specifically, we analyze the performance of the global and local measures under diverse structural configurations that encompass a broad range of location and scale heterogeneity among populations.
}

\subsection{Simulation results} \label{simures}
For illustrative purposes, the observations in this section were generated from log-normal distributions. In each scenario, four random variables, $X_1, X_2, X_3, X_4$, were simultaneously drawn from log-normal distributions to represent the $(m = 4)$ populations. To ensure consistency across sample sizes and minimize the impact of sampling variability on the asymptotic convergence analysis, we first constructed a master dataset comprising $n=500{,}000$ observations. Smaller sample sizes, ranging from $(1{,}000)$ to $(250{,}000)$, were then obtained by subsampling from this master dataset in such a way that all observations in smaller samples were nested within those of the larger samples.

\vspace{-0.6cm}
\begin{figure}[H]
$$
\begin{array}{cc}
\includegraphics[width=0.47\linewidth]{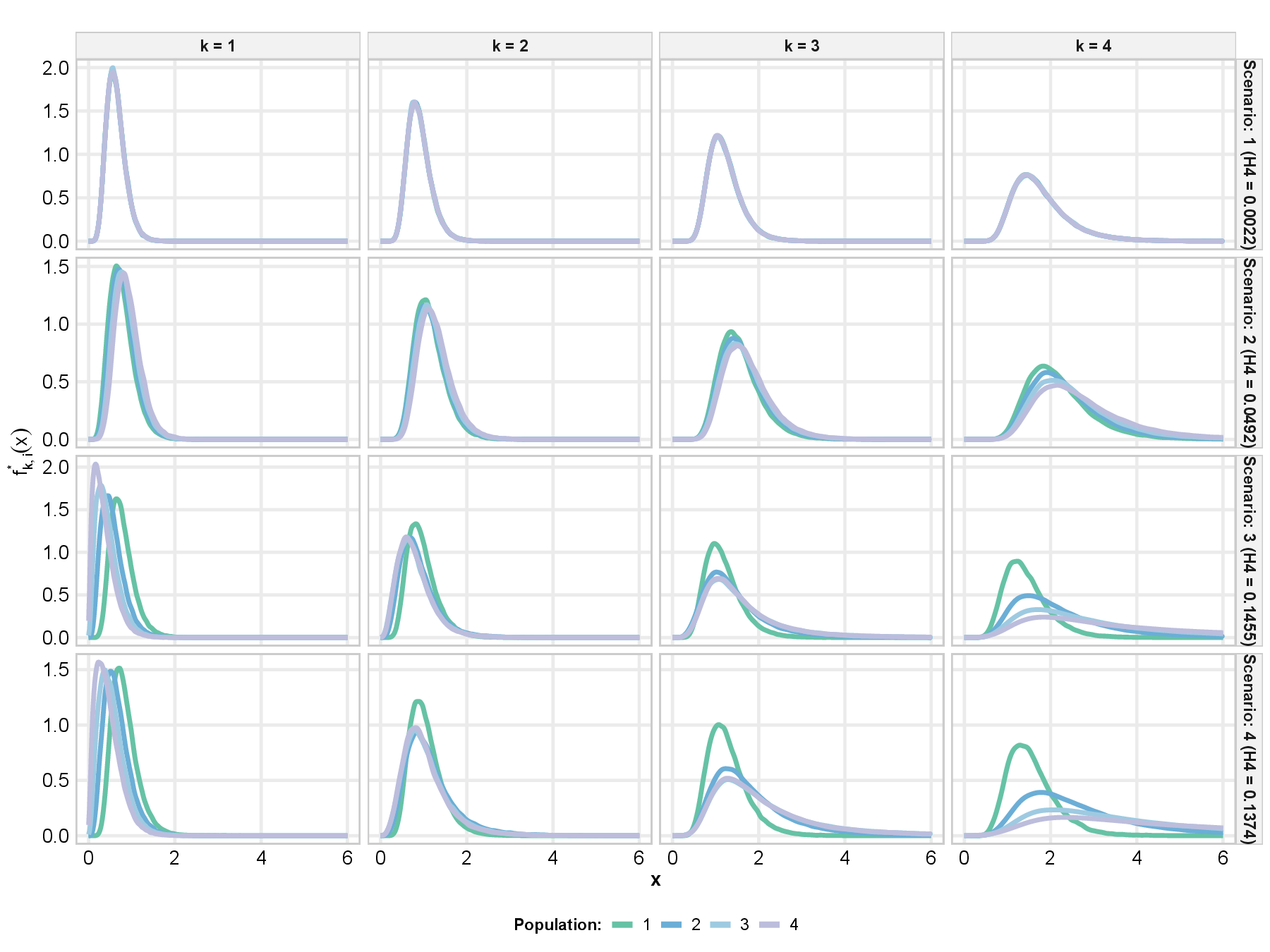}& \includegraphics[width=0.47\linewidth]{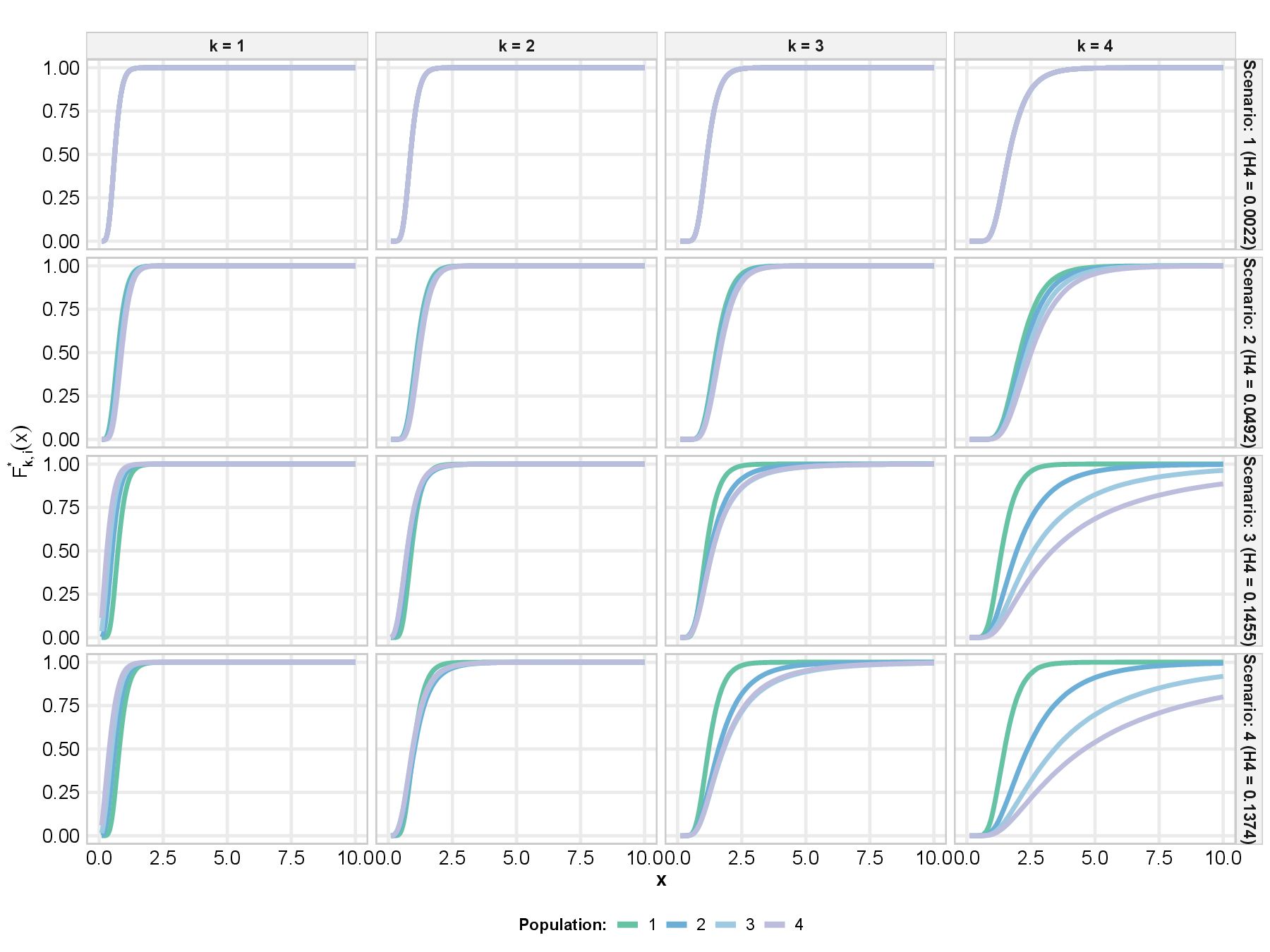}\\
\end{array}
$$
\vspace{-0.9cm}
\caption{Conditional density, $f_{k,i}^{*}(x)$, and cumulative distribution, $F_{k,i}^{*}(x)$, functions across ranks (\(k=1,\ldots,4\)) and simulation scenarios, along with the corresponding estimates of $H_4$.}
\label{Fig_SimuRes}
\end{figure}

\begin{figure}[H]
\centering
\begin{threeparttable}
\includegraphics[width=0.47\linewidth]{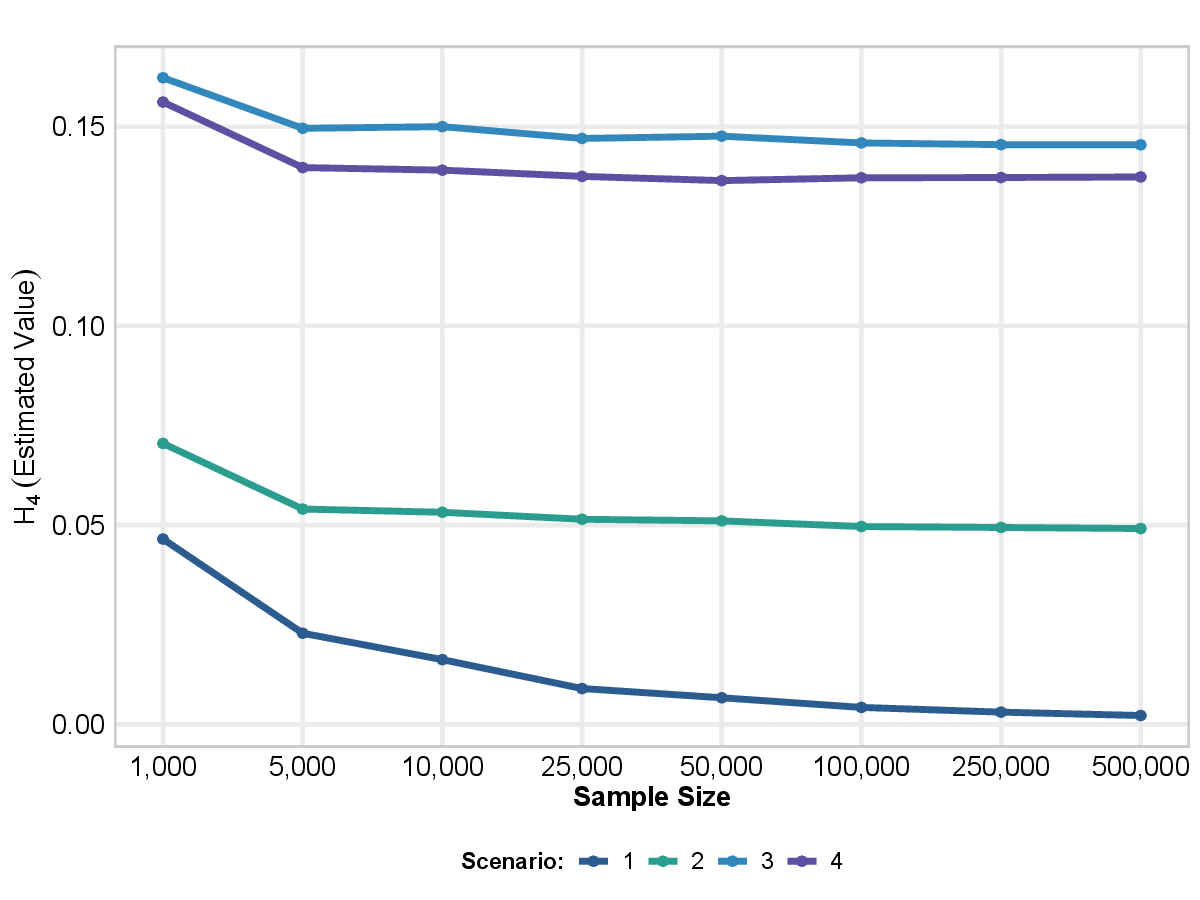}
\end{threeparttable}
\vspace{-0.9cm}
\caption{Asymptotic convergence of the global heterogeneity index $H_4$ across sample sizes under the four simulation scenarios is demonstrated.}
\label{Fig_AsympConvergence}
\end{figure}

The study considers four distinct structural scenarios to evaluate the effects of varying forms and degrees of heterogeneity among populations. In the first scenario (Independence), the four random variables are independent and identically distributed, with $\mu_j = 0$ and $\sigma_j = 0.5$ for all $j=1,2,3,4$, resulting in $H_3\approx 0$. The second scenario (Position) introduces progressive differences exclusively in the location parameter, $\mu_j$, while keeping the scale parameter fixed $(\sigma_j=0.5)$; specifically, $\mu_{j+1}=\mu_j+0.2$ with $\mu_1 = 0$. The third scenario (Scale) varies only the scale parameter, holding the location parameter constant, so that $\mu_j=0$ and $\sigma_{j+1} = \sigma_j+0.3$ with $\sigma_1=0.4$. Finally, the fourth scenario (Position and Scale) represents a more complex setting in which both location and scale parameters vary simultaneously across populations, allowing for the investigation of the proposed procedure under combined sources of distributional heterogeneity: $\mu_{j+1}=\mu_j+0.2$ and $\sigma_{j+1}=\sigma_j+0.3$ for all $j=1,2,3,4$.

Figures~\ref{Fig_SimuRes} and~\ref{Fig_AsympConvergence} illustrate the statistical stability and asymptotic convergence of the global estimator $H_4$ with increasing sample size (n). Specifically, $H_4$ approaches zero under the independence scenario (Scenario 1), whereas it converges to nonzero values reflecting the degree of structural divergence in the alternative scenarios (Scenarios 2--4). Moreover, the conditional density and cumulative distribution functions ($f_{k,i}^{*}$ and $F_{k,i}^{*}$, respectively) exhibit strong discriminatory power across ranks. The effects of location, scale, and their combination are clearly reflected in the corresponding curves, particularly at higher ranks ($k = 3$ and $k = 4$).

\subsection{Application results} \label{ApLres}

The \citet{hoel1972} dataset comprises two groups: Control ($n = 99$) and Germ-free ($n = 82$). The three causes of failure ($m = 3$) were treated as competing populations. In the Control group, the proportions were 22.22\% ($n = 22$) for thymic lymphoma, 38.38\% ($n = 38$) for reticulum cell sarcoma, and 39.39\% ($n = 39$) for other causes. In the Germ-free group, the corresponding proportions were 35.37\% ($n = 29$), 18.29\% ($n = 15$), and 46.34\% ($n = 38$), respectively. The analysis was stratified by experimental group, with rank-conditioned survival times modeled using the log-normal distribution and parameters estimated by maximum likelihood. This stratification prevents potential distortions in $H_3$ that could arise from aggregating groups with different survival time scales, while enabling the assessment of heterogeneity in failure-time distributions across the three competing causes within each environmental condition.

\vspace{-0.5cm}
\begin{figure}[H]
$$
\begin{array}{cc}
\includegraphics[width=0.47\linewidth]{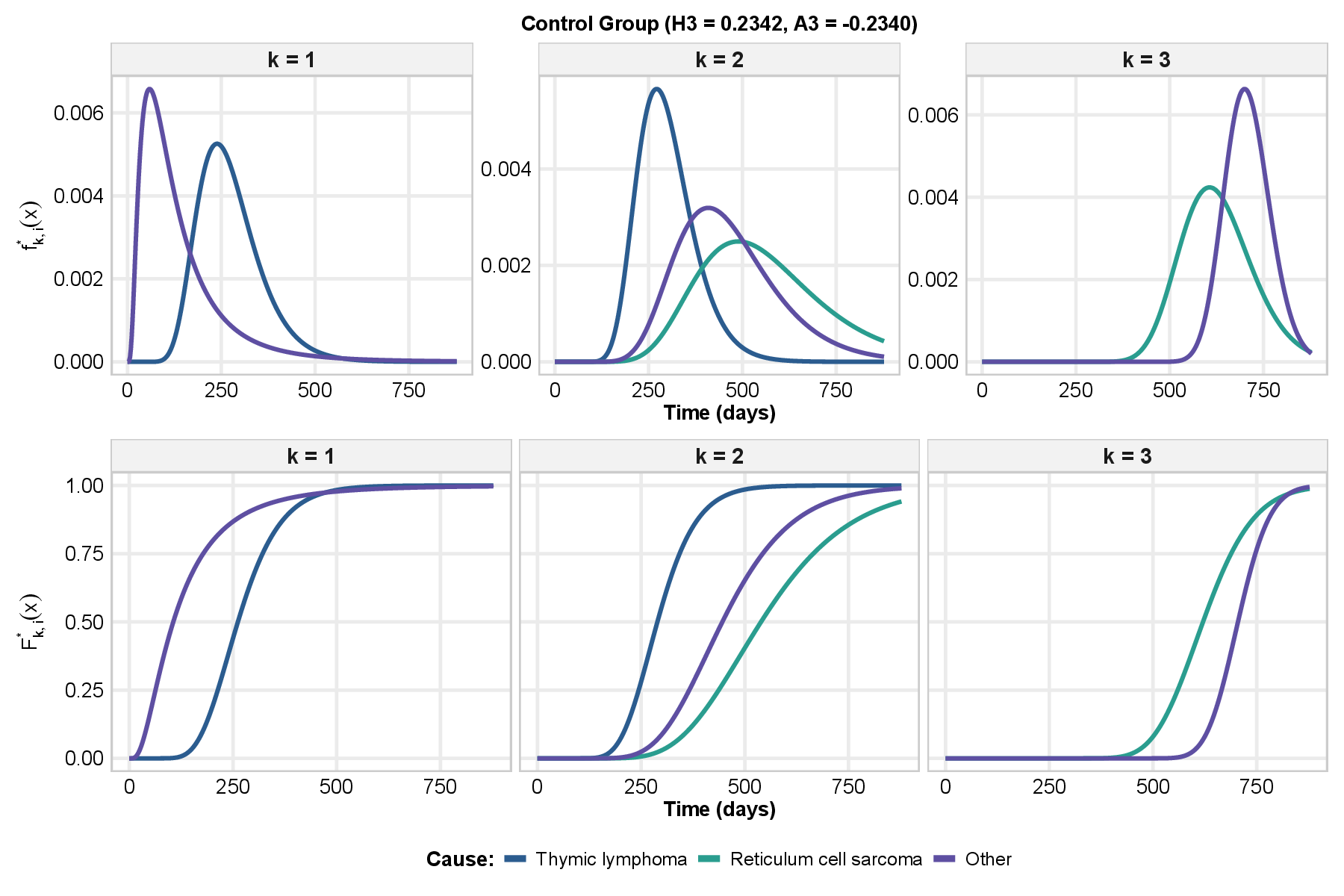}& \includegraphics[width=0.47\linewidth]{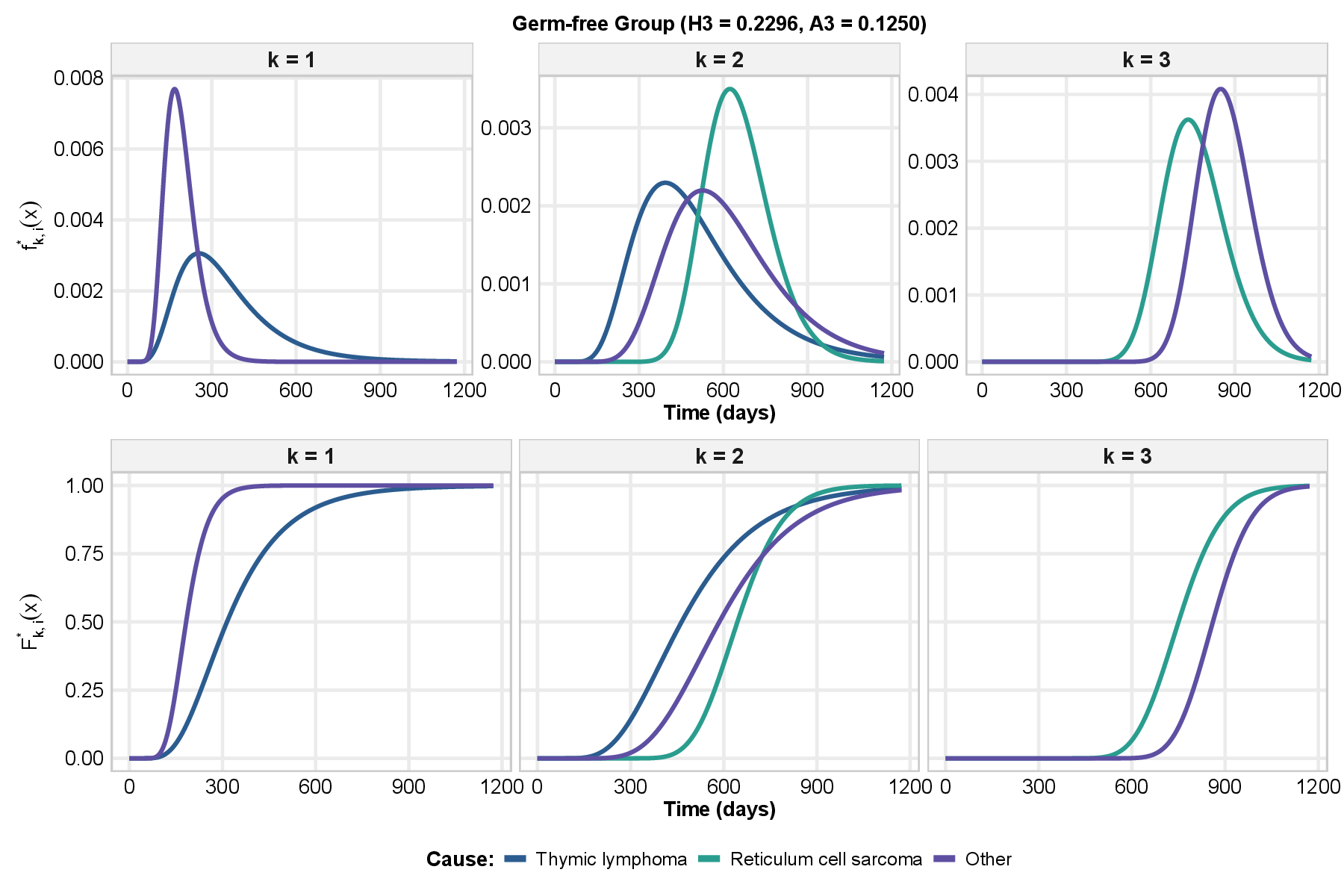}\\
\end{array}
$$
\vspace{-0.9cm}
\caption{Rank-conditioned functions for competing risks data ($m = 3$ causes), stratified by group (Control and Germ-free). The left and right panels display the functions for the Control and Germ-free groups, respectively.}
\label{FigAPL}
\end{figure}

The comparative stratified analysis of the Hoel (1972) data (see Figure~\ref{FigAPL}) demonstrates that the germ-free environment not only extends animal longevity but also alters the competitive dynamics among risks. The global heterogeneity measures indicate similar structural levels between groups (Control: $H_3 = 0.2342$; Germ-free: $H_3 = 0.2296$), reflecting comparable overall dispersion across failure cause ranks. In contrast, the asymmetry measure exhibits a pronounced sign inversion (Control: $A_3 = -0.2340$; Germ-free: $A_3 = 0.1250$), suggesting that the absence of microbiota shifts heterogeneity from early-stage failures to higher ranks and reduces the rigidity of temporal separation between pathologies. Notably, for the extreme ranks ($k = 1$ and $k = 3$), only two curves are shown instead of three, due to insufficient occurrences ($n_{ki} < 2$, where $n_{ki}$ denotes the number of observations experiencing cause $i$ at rank $k$) for certain causes to serve as the absolute minimum or maximum failure time.

Consequently, rank-conditioned log-normal parametric modeling demonstrates that the absence of microbiota systematically alters the hierarchy of competing causes. The environmental factor serves both as a global retarding mechanism for failure time and as a modulator of temporal dispersion throughout the organisms’ lifespan.

\section{Discussion and concluding remarks}
\label{Discussion}

This paper introduced a rank-associated measure of population
heterogeneity based on the conditional distributions of order
statistics. Unlike conventional measures of marginal
heterogeneity, the proposed measure accounts for the
association between population identity and rank in the pooled
sample. In particular, while $M_m$ compares the original
population distributions, $H_m$ quantifies the discrepancies
between the conditional and unconditional distributions of
the order statistics, thereby capturing heterogeneity
associated with pooled rankings.

The construction is related to the measures of departure from
independence studied in \cite{VilaAlmeida2026}, which focus on
the minimum and maximum in right- and left-censoring settings,
respectively. The present approach extends this perspective
to all order statistics and population indicators, providing
an aggregate assessment of rank-associated heterogeneity.
We established that $H_m=0$ if and only if each order
statistic is independent of its corresponding population
indicator $\delta_{k,i}$ whenever $p_{k,i}>0$. This
characterization connects the proposed measure to the
literature on independence and characterizations involving
order statistics and their associated indicators
\citep{BapatKochar1993,Malov1996,Kella(2024),VilaAlmeida2026}.
We also established boundedness, invariance under common
strictly monotone transformations, and bounds relating
$H_m$ to weighted pairwise distributional distances and
the diversity of population identities across ranks.

The use of the Kolmogorov--Smirnov distance allows
nonparametric comparisons of entire distribution functions
without requiring finite moments
\citep{Massey1951,GibbsSu2002}. Moreover, the probabilities
$p_{k,i}$ provide natural weights and determine an upper
bound for $H_m$ through the diversity of population
identities across ranks. The decomposition into minimum,
maximum, and intermediate contributions also helps identify
where rank-associated heterogeneity arises. For example,
in the exponential case, the minimum contribution vanishes,
whereas the maximum contribution is positive when the
populations have different rates. Thus, heterogeneity may
vary substantially across order statistics.

The numerical results reveal differences in rank-associated
heterogeneity even when marginal heterogeneity is identical.
Monte Carlo simulations support the stability and asymptotic
convergence of the proposed measure, while the empirical
application illustrates its ability to identify structural
differences in competing risks. In particular, the results
indicate a shift in failure heterogeneity toward higher ranks
in the absence of microbiota, a pattern that may not be
apparent from marginal distributional comparisons alone.
These findings highlight the complementary information
provided by incorporating population identity and rank
into the assessment of heterogeneity.

Some limitations of the proposed measure should be
acknowledged. The Kolmogorov--Smirnov distance captures
the largest absolute discrepancy between distribution
functions and may therefore overlook differences in moments
or absolute scale. Additionally, probability weighting can
reduce the influence of population--rank combinations with
small probabilities. Consequently, $H_m$ is best viewed as
a complementary measure of rank-associated population
heterogeneity rather than a universal measure of
distributional differences.

Future research could investigate alternative probability
distances, particularly Wasserstein-based measures, as
considered in \cite{VilaAlmeida2026} for the minimum and
maximum. Such alternatives may provide complementary
perspectives on distributional discrepancies, with different
sensitivities to location, scale, and tail behavior. Their
properties and relationships with marginal heterogeneity
and population diversity across ranks remain open questions
\citep{GibbsSu2002,SzekelyRizzo2013}.



\section*{Declarations}

\paragraph*{Ethics Approval}
Not applicable.



\paragraph*{Disclosure statement}
There are no conflicts of interest to disclose.

\paragraph*{Author Contributions Statement}
All authors contributed equally to the conception and design of the study, data analysis, interpretation of the results, manuscript preparation, and revision. All authors read and approved the final version of the manuscript.

\paragraph*{Data availability}

The dataset analyzed in this study is publicly available at \url{https://raw.githubusercontent.com/IowaBiostat/data-sets/main/Hoel1972/Hoel1972}, as described in \citet{hoel1972}.

{

}


\end{document}